\documentclass[sigconf,nonacm]{acmart}
\AtBeginDocument{%
  \providecommand\BibTeX{{%
    \normalfont B\kern-0.5em{\scshape i\kern-0.25em b}\kern-0.8em\TeX}}}

\usepackage{subcaption}

\setcopyright{none}
\copyrightyear{2024}
\acmYear{2024}
\acmDOI{XXXXXXX.XXXXXXX}
\acmConference[TREW '24]{}{May 11,
  2024}{Hawaii, USA}

\acmBooktitle{TREW ' 24: Trust and Reliance in Evolving Human-AI Workflows Workshop at CHI 2024,
 May 11, 2024, Hawaii, USA}

\begin{document}

\captionsetup[subfigure]{subrefformat=simple,labelformat=simple}
\renewcommand\thesubfigure{(\alph{subfigure})}

\newcommand{\centeredtxt}[1]{
    \begin{tabular}{l}
        \parbox{2cm}{\vspace{-20pt} \centering #1}
    \end{tabular}
}

\title{Facilitating Human-LLM Collaboration through Factuality Scores and Source Attributions}

\author{Hyo Jin Do}
\authornote{Both authors contributed equally to this research.}
\email{hjdo@ibm.com}
\affiliation{%
  \institution{IBM Research}
  % \streetaddress{P.O. Box 1212}
  \country{USA}
}

\author{Rachel Ostrand}
\authornotemark[1]
\email{rachel.ostrand@ibm.com}
\affiliation{%
  \institution{IBM Research}
  \country{USA}
}

\author{Justin D. Weisz}
\email{jweisz@us.ibm.com}
\affiliation{%
  \institution{IBM Research}
  \country{USA}
}

\author{Casey Dugan}
\email{cadugan@us.ibm.com}
\affiliation{%
  \institution{IBM Research}
  \country{USA}
}

\author{Prasanna Sattigeri}
\email{psattig@us.ibm.com}
\affiliation{%
  \institution{IBM Research}
  \country{USA}
}

\author{Dennis Wei}
\email{dwei@us.ibm.com}
\affiliation{%
  \institution{IBM Research}
  \country{USA}
}

\author{Keerthiram Murugesan}
\email{keerthiram.murugesan@ibm.com}
\affiliation{%
  \institution{IBM Research}
  \country{USA}
}

\author{Werner Geyer}
\email{werner.geyer@us.ibm.com}
\affiliation{%
  \institution{IBM Research}
  \country{USA}
}

% \author{}
%\renewcommand{\shortauthors}{Anon.}
\renewcommand{\shortauthors}{Do and Ostrand, et al.}

\begin{abstract}
While humans increasingly rely on large language models (LLMs), they are susceptible to generating inaccurate or false information, also known as ``hallucinations''. Technical advancements have been made in algorithms that detect hallucinated content by assessing the factuality of the model's responses and attributing sections of those responses to specific source documents. However, there is limited research on how to effectively \textit{communicate} this information to users in ways that will help them appropriately calibrate their trust toward LLMs. To address this issue, we conducted a scenario-based study (N=104) to systematically compare the impact of various design strategies for communicating factuality and source attribution on participants' ratings of trust, preferences, and ease in validating response accuracy. Our findings reveal that participants preferred a design in which phrases within a response were color-coded based on the computed factuality scores the most. 
Participants found it easy to validate the accuracy of an LLM's response and increased their trust in this style compared to a baseline in which no style was applied.
Additionally, participants increased their trust ratings when relevant sections of the source material were highlighted or responses were annotated with reference numbers corresponding to those sources, compared to when they received no annotation in the source material. Our study offers practical design guidelines to facilitate human-LLM collaboration and it promotes a new human role to carefully evaluate and take responsibility for their use of LLM outputs.
\end{abstract}


\maketitle
\section{Introduction}
% LLMs are popular but hallucinate
The rapid advancement of natural language generation technologies has led to the widespread use of large language models (LLMs) such as GPT, Bard, and LLaMA, in various tasks and contexts. However, LLMs are prone to presenting factually incorrect information as if it were true, a phenomenon known as ``\textit{hallucination}"~\cite{ji2023survey}. The presence of these hallucinations in LLM outputs, coupled with users' inability to easily detect them and the tendency to over-trust LLMs, has resulted in several high-profile incidents. These include lawyers being reprimanded by judges for presenting nonexistent case law that had been hallucinated~\cite{Sloan_2023}, new products being rapidly shelved due to hallucinated scientific references~\cite{ryan22meta}, news outlets issuing corrections to articles written with AI assistance~\cite{sato23cnet}, and company share prices falling after a hallucination leads to a blunder during a new product demo~\cite{Google_Webb}. 
Upon realizing the hallucinations, users may lose trust in LLMs, which further impedes technology adoption~\cite{wu2011meta}.
Researchers are actively investigating methods to mitigate hallucinations, such as refining and improving datasets and models~\cite{ji2023survey} using techniques such as reinforcement learning with human feedback~\cite{ouyang2022training, ji2023survey} and retrieval-augmented generation~\cite{lewis2020retrieval, cai2022recent}. However, technical advancements alone cannot completely resolve the issue; ultimately, it falls upon end-users to build an appropriate level of trust, be trained in how to carefully evaluate LLM outputs, and be accountable for their use.

Human-centered evaluation approaches offer a promising solution to address the hallucination issue. Several new methods have been developed that aid users in assessing the factual accuracy of a model's response. Notably, these include \textit{factuality scoring}, which evaluates the extent to which a model's response is truthful to a source document ~\cite{kryscinski2020evaluating,laban2022summac,maynez2020faithfulness,zhou2023synthetic, chern2023factool,min2023factscore} and \textit{source attribution}, which links the generated response to its source material~\cite{nakano2021webgpt, li2023improving, syed2021summary, evidence_inspector, perplexity}. 
However, it is currently unclear how to effectively \textit{communicate} this factuality information to users. 
Should it be conveyed numerically or visually? At what level of linguistic granularity should such information be presented (e.g. word, phrase)? 
Recently, \citet{leiser2023chatgpt} conducted participatory workshops where people brainstormed design strategies to identify hallucinations in the LLM outputs. However, no studies have systematically compared the effectiveness of these strategies in helping users assess the accuracy of the model's response and calibrate their trust. Our study aims to identify the most effective and preferred strategy for communicating two pieces of information about an LLM's response: (1) the \textit{factuality score}: the automated assessment of how factual the response is, and (2) \textit{source attribution}: identification of the sections within a source document from which the response was generated.

We explore various design strategies for communicating factuality information in a question-and-answer scenario. We first developed a factuality score color scale ranging from 0 (red) to 1 (green). We presented participants with three styles for visualizing factuality scores within an LLM's response: (1) \textit{highlight-all}, which annotates all of the linguistic content in the LLM response with varying background colors according to the scale, (2) \textit{highlight-threshold}, which annotates only those parts of the LLM response where the factuality score is below a given threshold, and (3) \textit{score}, which shows the numeric factuality score associated with each part of the response. Factuality scores were evaluated at two levels of linguistic granularity -- \textit{phrase} and \textit{word} -- and the three factuality styles were presented at each level of granularity. We additionally investigated two styles for presenting source attribution: (1) \textit{highlight gradients}, in which linguistic components of the source that were used to generate the model's response are highlighted, and (2) \textit{reference numbers}, which displays in-line citations within the model's response to specific, numbered parts of the source.

We conducted a scenario-based survey study (N=104) to compare the effects of these design strategies on participants' ratings of trust, preference, and ease of evaluating response accuracy. 
Based on our findings, this paper makes three contributions to the literature on human-AI collaboration: 1) We explore the design space for presenting factuality and source attribution information to users and identify a set of promising approaches for deep evaluation based on user feedback; 2) We show that our design strategies have significant effects on ratings of trust and ease of accuracy validation; 3) We offer practical guidance on how to communicate factuality scores and source attribution within the user interface of LLM-based applications and thereby facilitate human-LLM interactions.

\section{Related Work}

\subsection{Calibrating End-User Trust for Human-AI Collaboration}
Successful human-AI collaboration requires a user to modulate their level of trust in concert with the true reliability of the AI system, a process known as trust calibration~\cite{lee2004trust,wischnewski2023measuring}. 
Miscalibrated trust can lead to overreliance -- by accepting an incorrect AI recommendation -- or underreliance -- by not accepting a correct AI recommendation~\cite{wischnewski2023measuring}. 
Researchers identified three main factors that influence trust: 1) AI-related factors of performance (e.g., reliability, failure rate) and attributes (e.g., anthropomorphism), 2) user-related factors such as ability (e.g., expertise, prior experiences) and characteristics (e.g., demographic), and 3) environment-related factors such as team collaboration (e.g., culture, communication, reputation) and tasks (e.g., task complexity and type)~\cite{hancock2011meta,kim2023humans}. 
Kim et al. argued that participants with limited domain expertise had difficulty in assessing the accuracy of AI-generated outputs~\cite{kim2023humans}. 
Thus, we might expect that a user's assessment of the accuracy of the model and their trust in the model would be related, due to the interrelationship between accuracy assessment and expertise, and separately, trust and expertise. In particular, these prior works suggest that a user's trust in AI may be \textit{contingent} on their accuracy assessment of the AI response.
To get a full picture of trust, it is also crucial to examine both general trustworthiness perceptions such as user expectations in the use of LLMs, and instance-specific trust-related behaviors such as users' accuracy assessments of the LLM-generated output in the case of specific model responses~\cite{kim2023humans}.
In our work, we measured users' ratings of the ease of validating the model response's accuracy, their trust in the model, and their overall preferences among the different designs of factuality. We provided empirical evidence of trust calibration depending on their initial accuracy assessment of the model's output.

\subsection{Hallucination and Factuality Detection in Large Language Models}
\label{sec:relwork:factuality}
The widespread usage of LLMs in society has drawn attention to their risks and limitations. Notably, LLMs have the potential to generate text that seems plausible at first glance, but in reality, it is factually incorrect, a phenomenon known as \textit{hallucination}. In the context of natural language generation technologies, hallucination refers to the generation of content that is factually inconsistent or unfaithful to the provided source. The counterpart of hallucination is \textit{factuality}, defined as ``truthfulness or the quality of being based on fact''~\cite{ji2023survey}. 
The \textit{source} document is essential in determining the factuality of an LLM output. If the model's response aligns with the information from a reliable source, it is likely to be factually correct. 
On the other hand, \textit{faithfulness} means the LLM-generated response stays consistent with the source. 
In this study, we assume a reliable source as our basis for ``fact" so that faithfulness has the same meaning as factuality~\cite{maynez2020faithfulness}. 

% Factuality
Hallucination in LLMs can stem from various factors, such as noisy, biased, and/or erroneous training data, as well as the model itself. As summarized in survey papers~\cite{ji2023survey, huang2023survey}, researchers have addressed data-related issues by establishing ground truth data through human annotators and enhancing model inputs with external knowledge~\cite{honovich2022true,ji2023survey, huang2023survey}. 
However, it is impossible to completely resolve the hallucination issue inherent in AI technologies. 
Ultimately, it is the responsibility of end-users to carefully evaluate and be accountable for the use of LLM responses. 
As part of the effort to assist end-users in evaluating LLM responses, there has been ongoing research to develop methods for scoring the factuality of LLM outputs ~\cite{laban2022summac,lundberg2017unified_SHAP, maynez2020faithfulness,zhou2023synthetic, chern2023factool}. These methods can either use lexical matching-based metrics~\cite{lin2004rouge,banerjee2005meteor,papineni2002bleu} or model-based metrics using neural networks~\cite{zhang2019bertscore, sellam2020bleurt, min2023factscore}.
This increasing body of research raises new questions for LLM developers and designers on how to effectively \textit{communicate} factuality information to end-users. 
Specifically, there are no guidelines on which parts of the LLM response (e.g., correct, incorrect, or both) should be annotated, in what visual style (e.g., numeric, color-coded), and at what level of linguistic granularity accuracy should be assessed (e.g., word, phrase, entire response).
Furthermore, we have little understanding of how communicating the factuality of LLM outputs mitigates the effects of hallucination and calibrates end-users' trust. 
In addressing this gap, the present work identifies the most preferred and effective design to communicate the factuality and source attribution of LLM outputs, and presents practical guidelines based on our findings.

\subsection{Source Attributions as Explanations}
Source attribution is connected to research on explainable AI (XAI), providing explanations to support appropriate human understanding of AI-generated outputs~\cite{ribeiro2016should,lundberg2017unified_SHAP,sundararajan2017axiomatic,liao2020questioning,chen2020generating,kim2020interpretation,yin2022interpreting,mosca2022shap,ju2023hierarchical}. 
Liao and Vaughan emphasized the importance of communicating and being transparent about information during interactions with LLMs, given that LLMs raise unique challenges in XAI including new and complex model capabilities, behaviors, and applications of LLMs, massive and opaque architectures, and organizational pressure to move fast and deploy at scale~\cite{liao2023ai}. 
We investigated two design strategies -- highlight gradients and reference numbers -- to explain which parts of a source are most relevant to the LLM-generated response.
These design strategies are widely used in LLMs~\cite{syed2021summary,ribeiro2016should,lundberg2017unified_SHAP,chen2020generating,yin2022interpreting,ju2023hierarchical,gao2023rarr,gao2023enabling} and other similar contexts that incorporate sources (e.g., news media~\cite{sundar1998effect} and academic research citations~\cite{hyland1999academic}). 
Despite the common use of both design strategies in real-world applications and research, there is no research we are aware of that investigates the effects of these strategies on end-users' trust and ease of validating the response accuracy within the context of LLMs, both of which we address in the present work.

\section{Methods}
The goal of the current study was to understand how to present information about factuality and source attributions to an end-user in a way that is easy to understand and helps them calibrate their trust in LLM-generated text. To achieve this goal, we first reviewed designs of commercial generative AI systems and prior research to understand how other researchers and designers have presented factuality and source attribution information, and we ideated additional ways to present this information to a user. For the controlled study, we selected six different designs for representing a factuality score and two different designs for representing source attributions. This selection was based on a pilot study in which we interviewed ten participants about their preferences on numerous design options.
%We approached this research question in a three-phase manner: (1) we conducted a design review of commercial generative AI systems and prior research to understand how other researchers and designers have presented factuality and source attribution information, and ideated additional ways to present this information to a user; (2) we conducted a pilot study to evaluate different design options and selected a subset of designs to evaluate; and (3) we conducted a controlled study to evaluate six different options for representing a factuality score and two different options for representing source attributions.

\subsection{Participants}
\label{sec:participants}

There were 104 participants in the controlled study, who were employees of a large, multinational technology company. %Our goal was to recruit a diverse participant sample in terms of geography, job role, language proficiency, and experience with AI, machine learning, and LLMs.
Participant recruitment was advertised widely within the company on 25 internal Slack channels spanning multiple divisions and geographic regions in order to recruit a diverse sample across multiple professional, demographic, and experiential characteristics. 
Participants' work locations consisted of 20 unique countries, and %, with the most common being the US (N=53), India (N=8), the UK (N=8), Brazil (N=5), and France (N=4). 
job roles spanned a wide array of disciplines, including design, customer service, engineering, sales, research, and HR, among others.
Participants had a range of experiences with AI as a technology, with some having heard about it from the news, work, friends or family (N=14), others reporting that they ``closely follow'' AI news (N=26), many reporting some work or educational experience regarding AI (N=49), and a small number with ``significant'' work experience with AI (N=15). Participants also reported a wide range of experience with LLMs, with daily usage (N=10), a few times a week (N=21), once a week (N=10), once a month (N=15), a couple of times a year (N=15), a couple of times in life (N=13), never (N=19), and not sure (N=1). %The experiment was conducted in English; participants also had varying degrees of English exposure and proficiency. 
Participants self-rated their English proficiency on a 7-point Likert scale, with 68\% rating themselves at \textit{7 (native or native-like proficiency)}, 19\% as \textit{6}, 8\% as \textit{5}, 4\% as \textit{4 (medium)}, and 1\% as \textit{3}.
All participants provided written informed consent and were treated in accordance with the guidelines for the ethical treatment of human participants.

\subsection{Procedure}
\label{sec:procedure}
Participants were told to put themselves in the place of a user of an AI-powered language model. Their task was to evaluate different designs for presenting the factuality of the model’s response, and the source that the response was drawn from. On each trial, participants were shown a snapshot of a supposed interaction with an LLM, with three components: a \textit{Question}, a \textit{Response}, and a \textit{Source}. 
The Question was ``What movies did Beyonce star in and with whom?'' and the AI-generated Response was ``Beyonce starred in the musical comedy The Fighting Temptations in 2002 and in the documentary film Austin Powers in Goldmember in 2003, alongside Missy Elliott and Foxxy Cleopatra, respectively.'' The source was a paragraph from Wikipedia, and we asked participants to assume that the source was factually accurate. The response was was written to have some factually inaccurate propositions (and importantly, contradictory of the source document) and other accurate ones; overall, it was approximately half accurate and half inaccurate.
The question and the response were carefully selected to test our design strategies, and covered a topic that was not technical, to enable non-expert participants to make their own assessment as to its accuracy.

During the study, participants were shown a series of different design strategies to evaluate. Each design was demonstrated using the same Question, Response, and Source text, to hold constant the content and level of accuracy across different designs. This allowed us to reduce the number of variables tested and ensure a more targeted exploration of our design strategies. Participants were always shown the \textit{Baseline} design first, which had no markup, and displayed only the text of the Question, Response, and Source. % (see Fig.~\ref{fig:baseline}). 
Next, participants were presented with six different design strategies for displaying the model response's factuality, and finally, two different design strategies for showing source attribution in the response.
The study was conducted as a within-subjects experiment, and thus all participants viewed and rated the same designs.

% \begin{figure}[ht]
% \includegraphics[width=.8\linewidth]{Figures/baseline-full.png}
% \caption{The baseline design was shown to participants at the start of the experiment, with no annotations showing either factuality information or source attribution.}
% \label{fig:baseline}
% \end{figure}

As a first step, participants were asked to rate their perceptions about the model and its response on a 7-point Likert scale for the \textit{Baseline} design along three dimensions:
\begin{enumerate}
    \item \textit{Perceived accuracy}: How accurate do you think this AI-generated response is?
    \item \textit{Ease of validation}: With the information presented in this way, how easy is it for you to determine the accuracy of this AI-generated response?
    \item \textit{Trust}: With the information presented in this way, how much do you trust the AI system that generated the response?
\end{enumerate}

\subsubsection{Factuality Score}
Following the baseline design, participants were introduced to the concept of a \textit{factuality score} – a feature that compares linguistic components of the response against the source – and that a high factuality score indicated that the response aligned with the information in the source and thus is likely to be correct. 
Three factuality design strategies were presented to each user, each at two levels of granularity. The designs were \textit{highlights-all}, in which every part of the response text was highlighted with a color-coding to show its level of factuality on a red (0) to green (1) scale; \textit{highlights-threshold}, in which only the sections of the response text with a factuality score below 0.5 were highlighted, to signal inaccuracies; and \textit{score}, in which all parts of the response text were tagged with their factuality score, but instead of highlights, with color-coded underlines and the numerical factuality score value.

% \begin{figure}[t]
%     \centering
%     \includegraphics[width=.4\linewidth]{Figures/scale.png}
%     \caption{Factuality scale. The color scale was presented to participants to demonstrate the range of factuality scores and their corresponding colors.} 
%     \label{fig:factuality_scale}
% %    \vspace{0.5cm}
% \end{figure}

\begin{table*}[h]
    \centering
    \caption{The set of designs presented to each participant for displaying factuality scores on the model's response. Each participant saw and rated all six designs, in a randomized order but grouped by granularity.}
    \begin{tabular}{cccc}
        Granularity & Highlight-all & Highlight-threshold & Score \\
        \centeredtxt{Word} & \includegraphics[width=0.25\textwidth]{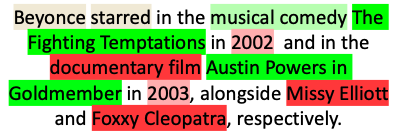} & \includegraphics[width=0.25\textwidth]{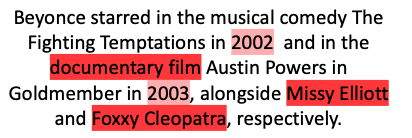} & \includegraphics[width=0.25\textwidth]{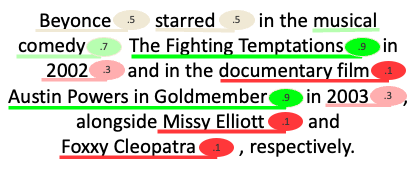} \\
        \centeredtxt{Phrase} & \includegraphics[width=0.25\textwidth]{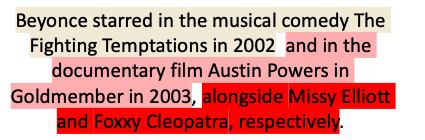} & \includegraphics[width=0.25\textwidth]{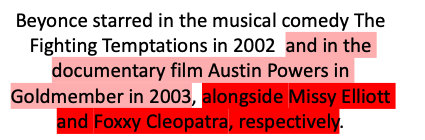} & \includegraphics[width=0.25\textwidth]{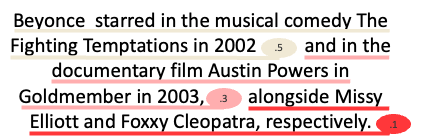} \\
    \end{tabular}
    \vspace{.3cm}
    \label{fig:factuality_designs}
\end{table*}

In addition, the designs were presented at two levels of \textit{granularity} – either \textit{word-level} or \textit{phrase-level}. This refers to the amount of text over which the factuality was evaluated. At phrase-level granularity, if there was an inaccuracy in one word in a phrase, then the entire phrase would be tagged with a lower factuality score. In contrast, at word-level granularity, only that word or a group of words would be tagged with a lower factuality score, while the other words in the sentence (if correct), would individually be tagged with a higher factuality score. Table~\ref{fig:factuality_designs} shows the six designs that users evaluated.

Participants were shown each factuality design strategy one at a time, and asked to rate their perceptions on two dimensions: ease of validation and trust (questions (2) and (3) as listed in Section~\ref{sec:procedure}), using a 7-point Likert scale.
Note that we did not ask users about perceived accuracy (question (1)) for any designs except for the Baseline, because the wording of the text was identical in the Baseline and every presented design.

Participants performed this rating task for the three designs at one granularity (word-level vs. phrase-level), and then were asked to rank-order them, along with the baseline, in their order of preference. They then performed the same rating and preference-ranking task for the three designs at the other granularity. The three designs within each granularity were presented in a randomized order to each participant, and the order of the two granularities was also randomized across participants, to reduce possible order effects in the aggregated data.
Following the ratings, we asked a supplemental question regarding preference between the two types of granularity designs. 

\subsubsection{Source Attribution}
\begin{figure}[ht]
 \begin{subfigure}{.48\textwidth}
    \centering
     \includegraphics[width=\linewidth]{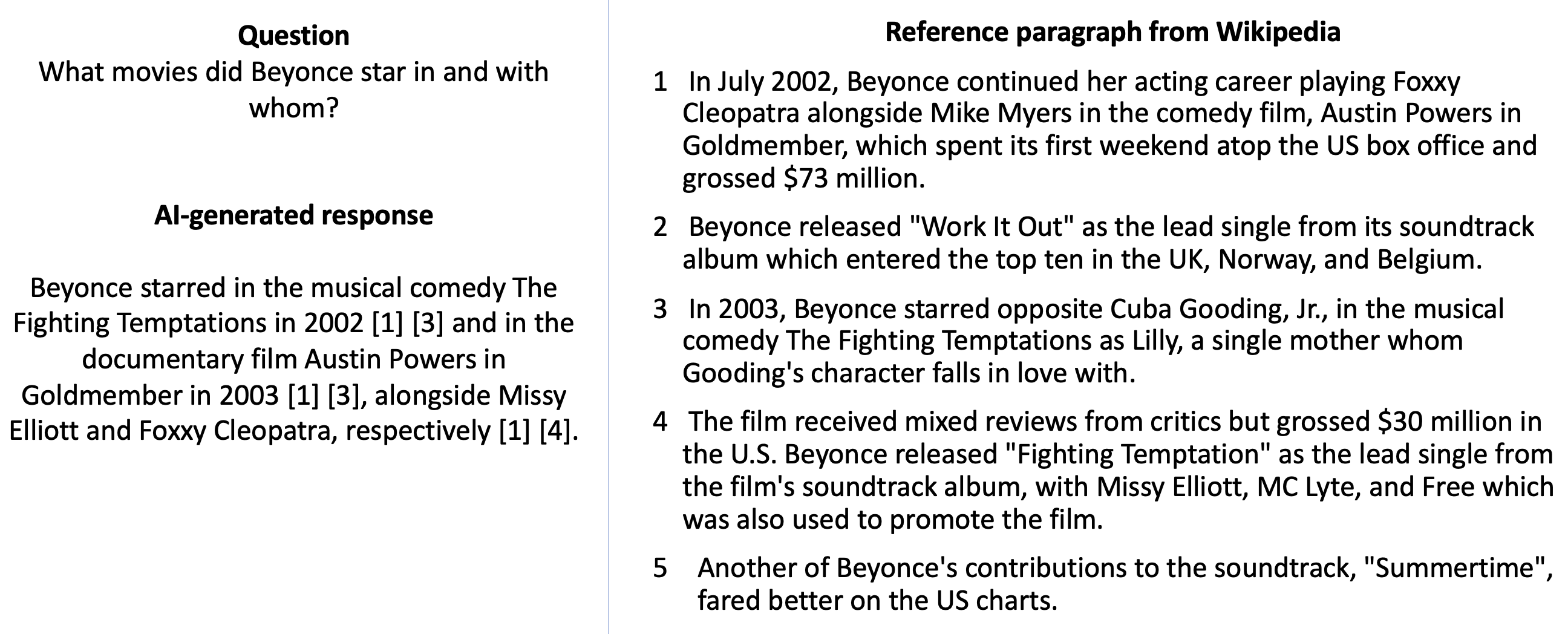}
     \caption{Reference numbers}
     \label{attribution-numbers}
     \vspace{.3cm}
 \end{subfigure}
 \begin{subfigure}{.48\textwidth}
    \centering
   \includegraphics[width=\linewidth]{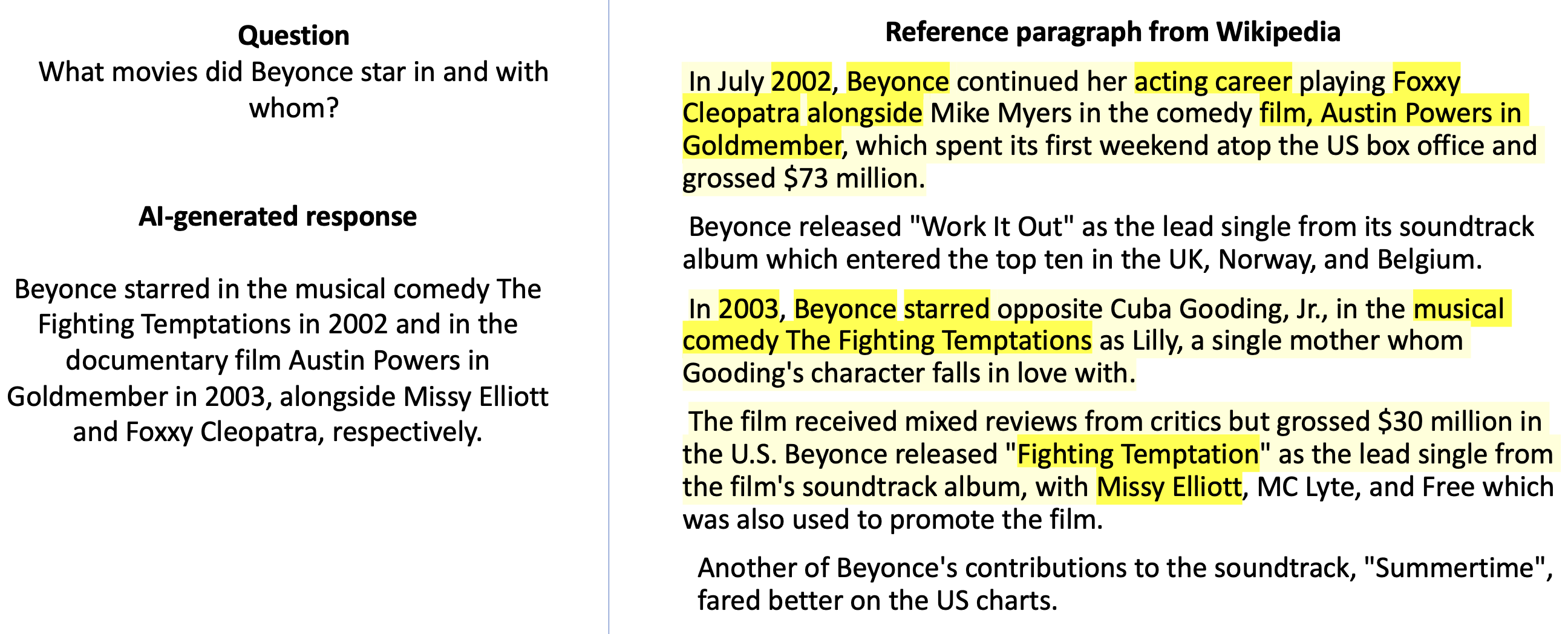}
   \caption{Highlight gradients}
   \label{attribution-gradients}
   \vspace{.3cm}
 \end{subfigure}

\caption{The set of designs presented to each participant for displaying the source attribution by the model. Each participant saw and rated both designs, in a randomized order.}
\label{fig:attribution_designs}
\end{figure}

Next, participants were introduced to the source attribution feature, in which the source was annotated to show which parts were used to generate the model's response. There were two designs that were presented to users: \textit{reference numbers}, in which each sentence of the source document was numbered, and propositions in the response were tagged with the number corresponding to the source sentence from which it was derived; and \textit{highlight gradients}, in which sections of the source that provided the information for parts of the response were highlighted. The order of presentation of the two source attribution designs was randomized between participants. Fig.~\ref{fig:attribution_designs} presents the two designs that users evaluated.

After each source attribution design, participants were asked the same two questions as for the factuality score designs.
After rating the two source attribution designs, participants were then asked to rank-order their preference among the two source attribution designs and the baseline. At the end of the survey, participants responded to some demographic and professional questions as reported in Section~\ref{sec:participants}.

\subsection{Participants}
\label{sec:participants}

There were 104 participants in the study, who were employees of a large, multinational technology company. 
As such, participant recruitment was advertised widely within the company on 25 internal Slack channels spanning multiple divisions and geographic regions. 
All participants provided written informed consent and were treated in accordance with the guidelines for the ethical treatment of human participants. 
Participants' work locations consisted of 20 unique countries. 
Job roles spanned a wide array of disciplines, including design, customer service, engineering, sales, research, HR, among others.
Participants had a range of experiences with AI as a technology, with some having heard about it from the news, work, friends or family (N=14), others reporting that they "closely follow" AI news (N=26), the largest subset reporting some work or educational experience regarding AI (N=49), and others with "significant" work experience with AI (N=15). Participants reported a wide range of experience with LLMs, and varying degrees of English exposure and proficiency.

\section{Results}
The analyses were conducted using generalized linear mixed-effects models, with one model for each dependent variable: participants' ratings of (1) trust and (2) ease of validating the response accuracy. In both models, the categorical independent variable Design Strategy was treatment-coded with the Baseline design set as the reference level, such that each design's rating was statistically compared against the baseline. Participant ID was a random variable. Following the omnibus models, pairwise contrasts were conducted to explore comparisons between each pair of Design Strategy levels, with \textit{p}-values adjusted for multiple comparisons using the Tukey correction.

\subsection{Factuality Score}
\subsubsection{Trust}

First, we compared users’ ratings of their trust of the model that produced the response, for each of the designs compared against the baseline. All six of the designs were rated as significantly more trustworthy compared to the baseline, suggesting that all of the designs that presented response factuality increased users' trust in the model. The mean and the standard error (SE) of the ratings for each design, along with results from the statistical model comparing each design to the baseline, are displayed in Table~\ref{table:factuality-trust}.
Post-hoc pairwise comparisons between all pairs of designs revealed no additional significant differences after correction for multiple comparisons.

\begin{table*}[h]
\caption{Users’ ratings of (a) their trust in the model and (b) the ease of assessing the accuracy of the model's response, for the baseline and each of the factuality score designs. \textit{t} and \textit{p} values were calculated within the omnibus statistical model, and represent the comparison of each Design Strategy against the Baseline as the reference level. Bolded Design Strategy names indicate a significant difference from the Baseline.}
  \begin{subtable}{.5\linewidth}
    \centering
    \begin{tabular}{ lcccl } 
         \toprule
        Design Strategy & Mean &  SE & \textit{t} & \textit{p}\\
        \midrule
        Baseline  & 2.63 & 0.17 & - & - \\
        Phrase-level granularity \\
            \hspace{3mm}\textbf{Highlights-all}        & 3.31 & 0.17 & 4.59 & < .001 \\
            \hspace{3mm}\textbf{Highlights-threshold}  & 3.22 & 0.16 & 4.00 & < .001 \\
            \hspace{3mm}\textbf{Score}                 & 3.21 & 0.17 & 3.93 & < .001 \\
        Word-level granularity \\
            \hspace{3mm}\textbf{Highlights-all}        & 3.62 & 0.18 & 6.57 & < .001 \\
            \hspace{3mm}\textbf{Highlights-threshold}  & 3.49 & 0.18 & 5.80 & < .001 \\
            \hspace{3mm}\textbf{Score}                 & 3.54 & 0.18 & 5.85 & < .001 \\
        \bottomrule
    \end{tabular}
    \caption{User trust of the model}
    \label{table:factuality-trust}
  \end{subtable}%
  \begin{subtable}{.5\linewidth}
    \centering
    \begin{tabular}{ lcccl } 
         \toprule
        Design Strategy & Mean &  SE & \textit{t} & \textit{p}\\
        \midrule
        Baseline  & 4.29 & 0.20 & - & - \\
        Phrase-level granularity \\
            \hspace{3mm}\textbf{Highlights-all}        & 4.74 & 0.17 & 2.24 & .03 \\
            \hspace{3mm}Highlights-threshold           & 4.53 & 0.17 & 1.19 & .23 \\
            \hspace{3mm}Score                          & 4.38 & 0.17 & 0.48 & .63 \\
        Word-level granularity \\
            \hspace{3mm}\textbf{Highlights-all}        & 4.72 & 0.19 & 2.10 & .04 \\
            \hspace{3mm}\textbf{Highlights-threshold}  & 4.88 & 0.16 & 2.96 & .003 \\
            \hspace{3mm}Score                          & 4.32 & 0.20 & 0.13 & .89 \\
        \bottomrule
    \end{tabular}
    \caption{Ease of assessing the accuracy of the response}
    \label{table:factuality-validation}
  \end{subtable}
\end{table*}

\begin{figure*}[h]
\centering
    \begin{subfigure}{.4\textwidth}
        \includegraphics[width=\linewidth]{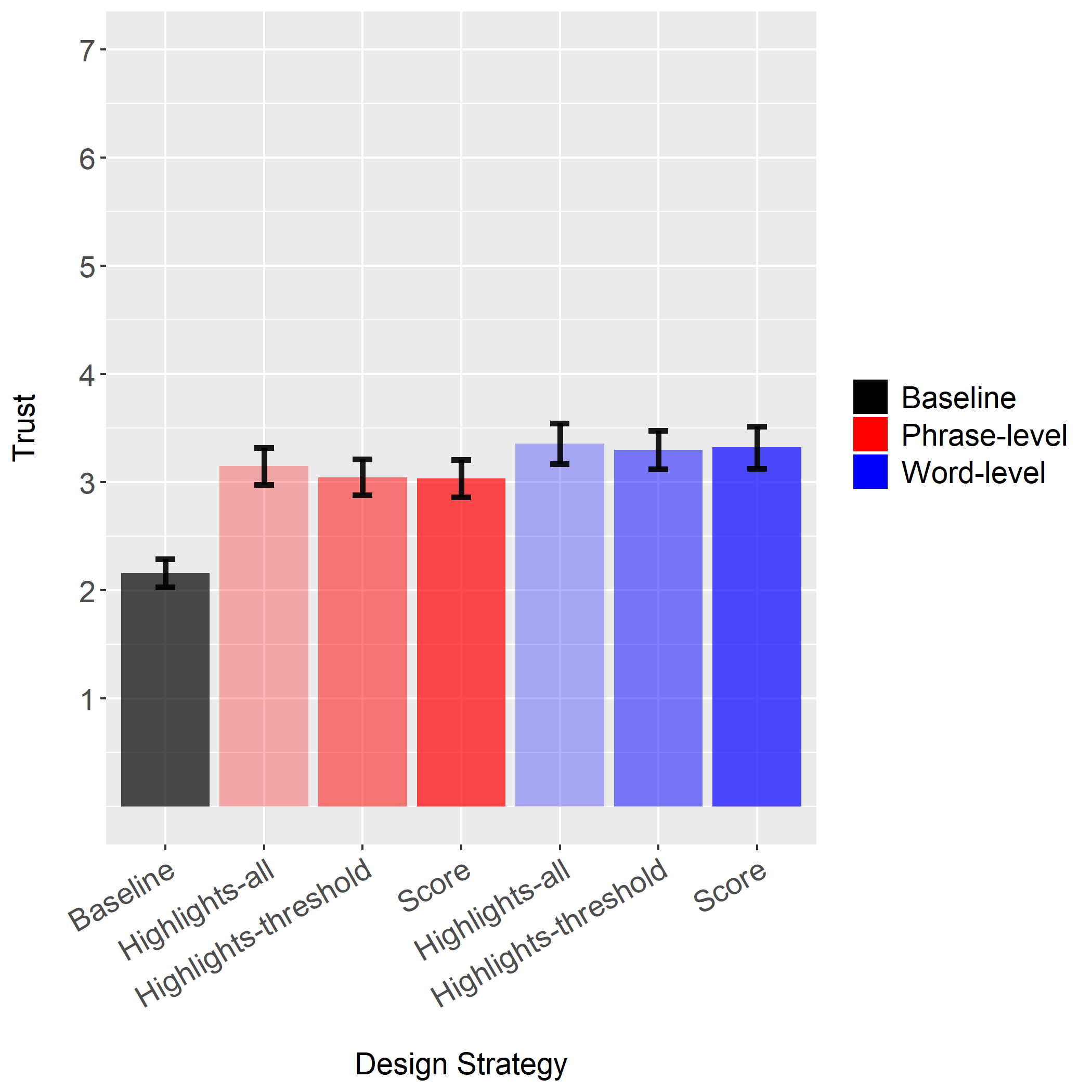}
        \caption{Participants with low baseline accuracy rating}
        \label{fig:low_baseline_acc}
    \end{subfigure}
    \hspace{10mm}
    \begin{subfigure}{.4\textwidth}
      \includegraphics[width=\linewidth]{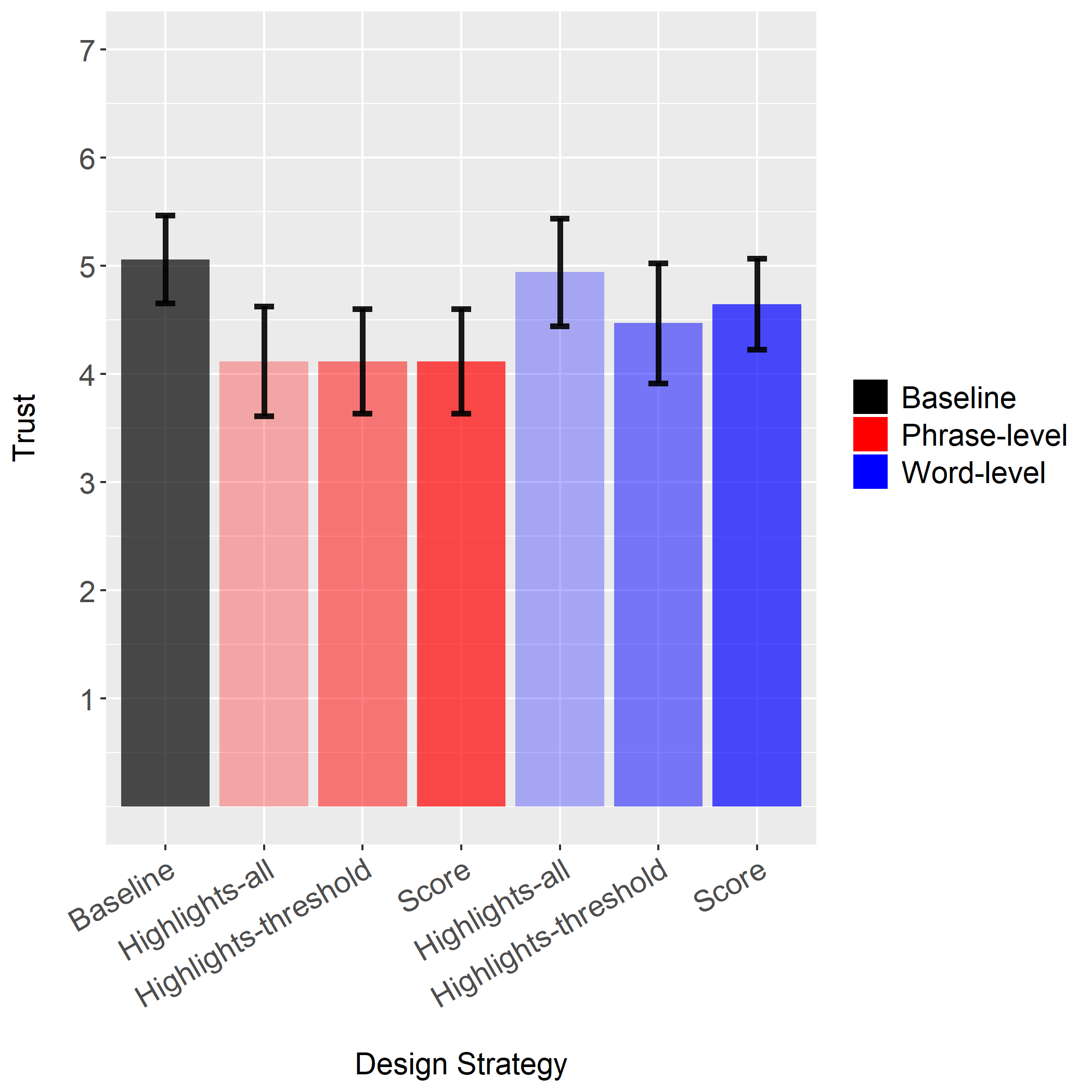}
      \caption{Participants with high baseline accuracy rating}
      \label{fig:high_baseline_acc}
    \end{subfigure}
\caption{Participants showed different levels of trust in the model as a function of their perceived accuracy at baseline. Participants who identified errors in the baseline response (a) reported higher levels of trust after reviewing factuality designs. Participants who initially missed errors in the baseline response (b) reported lower levels of trust after reviewing the factuality designs.}
\label{fig:factuality_trust_by_baseline_accuracy}
\end{figure*}

As an exploratory analysis, we investigated whether participants' baseline rating of the model response's accuracy affected how much they trusted the model when it subsequently provided factuality scores. To do so, we ran another linear mixed-effects model with the same structure as above, with the addition of \textit{perceived accuracy of the response at Baseline} as an independent variable.
Perceived baseline accuracy significantly affected participants' subsequent ratings of trust (\textit{t}~=~4.00, \textit{p}~<~.001). For visualization purposes, we categorized participants into two groups:
The \textit{low baseline accuracy} group, which was defined as those participants who rated the baseline response accuracy at or below 4 (N=87), and the \textit{high baseline accuracy} group, who rated the baseline response accuracy as 5 or higher (N=17). 
As illustrated in Fig.~\ref{fig:low_baseline_acc}, participants in the low baseline accuracy group initially rated the baseline with low trust, %as they identified many errors in the response. However, they 
but subsequently increased their trust ratings after reviewing the factuality scores. 
In contrast, participants in the high baseline accuracy group (Fig.~\ref{fig:high_baseline_acc}) initially had higher trust in the model's response but subsequently decreased their trust ratings after examining the factuality information presented in each design style, particularly at the phrase-level. 

\subsubsection{Ease of validation}
We next compared users’ ratings about the ease of assessing the model's accuracy for each of the designs, compared against the baseline. Of the six design strategies, three were rated as significantly easier to assess the response accuracy compared to the baseline: \textit{highlights-all} at phrase-level granularity, and \textit{highlights-all} and \textit{highlights-threshold} at word-level granularity. The other three designs were not rated significantly different from the baseline. Post-hoc pairwise comparisons between each pair of design strategies revealed no significant differences after correction for multiple comparisons.
The mean and standard errors of the ratings for each design, along with results from the statistical model comparing each design to the baseline, are displayed in Table~\ref{table:factuality-validation}.

\subsubsection{Preference}
Participants next rank-ordered each of the three designs plus the baseline, within each granularity level. Thus, each ranking compared four designs. The results are presented with the ranking scores reversed (i.e., 4 - ranking score) such that a higher score corresponds to a more preferred design for comparability with the trust and validation ratings. At phrase-level granularity, the \textit{highlights-all} design was the most preferred (M = 1.94), \textit{score} was second (M = 1.69), \textit{highlights-threshold} was third (M = 1.67) and baseline was the least preferred (M = 0.69). At word-level granularity, rankings were similar: the \textit{highlights-all} design was again the most preferred (M = 1.87), \textit{highlights-threshold} was second (M = 1.79), \textit{score} was third (M = 1.56), and the baseline was the least preferred (M = 0.78).

Participants were also asked their preference between the two types of granularities: 52.9\% of participants preferred \textit{phrase-level} granularity, while 26.9\% of the participants preferred \textit{word-level} granularity, with 10.6\% of the participants responding with ``don't know'' and 9.6\% of the participants selecting ``other''.

\subsection{Source Attribution}

\subsubsection{Trust}
Both \textit{reference numbers} and \textit{highlight gradients} caused the model to be rated as significantly more trustworthy compared to the baseline, as shown in Table~\ref{tab:attribution_trust_results}. Post-hoc pairwise comparisons between the two design strategies revealed no additional significant differences. 

\subsubsection{Ease of validation}
Both \textit{reference numbers} and \textit{highlight gradients} were rated significantly \textit{lower} (i.e., worse) compared to the baseline, as shown in Table~\ref{tab:attribution_accuracy_results}. Post-hoc pairwise comparisons between the two design strategies showed no significant differences between them. 

\subsubsection{Preference}
Participants were asked to rank-order their preference for the two designs and the baseline as a way to present source attribution information. As with the factuality scores, the data here are presented with the ranking scores reversed (i.e., 3 - ranking score) such that a higher score corresponds to a more preferred design. Participants preferred both of the two designs over the baseline, with \textit{reference numbers} the most preferred by a small margin (M = 1.21), followed by \textit{highlight gradient} (M = 1.17), and the least preferred was the baseline (M = 0.62).

\begin{table*}
\caption{Users’ ratings of (a) their trust in the model and (b) the ease of assessing the accuracy of the model's response, for the baseline and each of the designs for source attribution. Bolded Design Strategy names indicate significant differences from baseline.}
  \begin{subtable}{.5\linewidth}
    \centering
    \begin{tabular}{ lcccl } 
         \toprule
        Design Strategy & Mean &  SE & \textit{t} & \textit{p}\\
        \midrule
        Baseline  & 2.63 & 0.17 & - & - \\
            \hspace{3mm}\textbf{Reference numbers}        & 3.40 & 0.18 & 4.71 & < .001 \\
            \hspace{3mm}\textbf{Highlight gradients}      & 3.23 & 0.18 & 3.74 & < .001 \\
        \bottomrule
    \end{tabular}
    \caption{User trust of the model}
    \label{tab:attribution_trust_results}
  \end{subtable}%
  \begin{subtable}{.5\linewidth}
    \centering
    \begin{tabular}{ lcccl } 
         \toprule
        Design Strategy & Mean &  SE & \textit{t} & \textit{p}\\
        \midrule
        Baseline  & 4.29 & 0.20 & - & - \\
            \hspace{3mm}\textbf{Reference numbers}        & 3.81 & 0.19 & -2.18 & .03 \\
            \hspace{3mm}\textbf{Highlight gradients}  & 3.81 & 0.19 & -2.19 & .03 \\
        \bottomrule
    \end{tabular}
    \caption{Ease of assessing the accuracy of the response}
    \label{tab:attribution_accuracy_results}
  \end{subtable}
\end{table*}
\section{Discussion}
In this study, we offer insight into ways of presenting information to an end-user during human-AI collaboration, allowing the user to assess the factuality of LLM responses that have the potential to be hallucinated. This user study presented multiple design strategies for displaying factuality and source attribution information from an LLM. Here, we abstract across the different results to present design recommendations and discuss limitations.

The most common granularity preference among participants for presenting factuality information within a model's response was \textit{phrase-level granularity}.
Among the phrase-level granularity styles, the \textit{highlights-all} style, which color-codes each of the phrases in the response with the color associated with the computed factuality score, was the most preferred. The statistical results showed that the \textit{highlights-all} style also led to significantly higher trust and was significantly easier to validate response accuracy than the baseline. Thus, we make the design recommendation to present factuality information using a design similar to our \textbf{highlights-all at phrase-level granularity} design.

For source attribution, we recommend adopting either the \textbf{reference numbers} or \textbf{highlight gradients} design to enhance trust and accommodate the preferences of end-users, as both design strategies were effective in increasing trust compared to the baseline. The finding supports existing XAI research that explanations can increase user trust. The procedural justice theory~\cite{lind1988social} suggests that people's trust is strongly impacted by procedural explanations and not only the outcome.
Our design strategies provided a procedural explanation of how the model's response was generated, even where the response was partially inaccurate.
In contrast, participants commented that they had difficulty assessing the accuracy of the model's response using the design strategies because they felt overwhelmed and distracted by them. 
Therefore, we recommend incorporating a feature that enables users to turn off or filter styles to reduce distractions, or even remove the styles if prioritizing ease of validation over building high trust.

How participants perceived the accuracy of the model's response in the baseline design had a substantial impact on their trust after they viewed the factuality scores.
Participants who initially rated the model accuracy as high, despite the presence of multiple errors in its response, decreased their trust upon seeing the errors called out through the factuality scores. In contrast, participants who initially rated the model accuracy as low increased their trust when they observed that the factuality scores accurately flagged those errors. 
Expectancy violations theory supports the finding that positive violations (i.e., when perceived performance exceeds rather than meets the expected level of performance) have a stronger positive effect on satisfaction, while negative violations produce a negative effect~\cite{burgoon2016application}. 
Therefore, to calibrate the level of end-users' trust, these results support incorporating factuality information in the LLM response.

While our study assumed that the algorithm generating factuality scores for the model's response is reliable, the algorithm itself may be imperfect or erroneous. 
A similar situation exists for source attribution. We assumed the existence of a reliable source attribution algorithm, but in practice, source attribution and AI model explanation are ongoing research topics, especially for generative language models that output text rather than numerical predictions. These issues are heightened when there are multiple source documents, some of which may be irrelevant or unreliable. 
Therefore, end-users should be aware of the limitations of these AI technologies, particularly in the context of LLMs, which may appear factual at face value. It is crucial for users to always verify the model's responses across multiple sources to ensure the reliability of information and prevent themselves from placing over-trust in LLMs.

The current study had a few limitations that should be areas of future research. It focused on a single question-and-answer task, as there were multiple interventions (i.e., design strategies) to compare within this task.
Additionally, the model's response and the assigned factuality scores were handcrafted, rather than generating an actual LLM response and scores from existing factuality or source attribution algorithms. This allowed us to easily create designs that effectively tested our research questions, but real-world LLM responses and algorithms may be different. 
While we made efforts to recruit participants with diverse backgrounds, skills, and locations, our recruitment was restricted to individuals within our company. Finally, it is important to note that our research did not aim to exhaustively explore all potential design strategies. Instead, our study should be viewed as a starting point, encouraging researchers to delve deeper into diverse design strategies and expand the discussion.

\section{Conclusion}
Large language models have known problems with so-called hallucinations. To address these challenges, a growing area of research is the development of algorithms to assess the model response's factuality and attribute it to sources. However, how to effectively communicate factuality and source attribution to end-users is an open question. In this study, we designed and compared six design strategies for communicating factuality scores and two design strategies for conveying source attribution. Conducting a scenario-based study through an online survey, we discovered that highlighting every phrase in the model's response based on the factuality score is the most preferred strategy and leads to higher trust than the baseline without any markups. Our findings also revealed that participants calibrated their trust in the model based on their initial accuracy assessment of the response. Regarding source attribution, reference numbers and highlight gradients enhanced trust in the model, but did not alleviate the challenge of assessing the response accuracy. We provide practical guidance on communicating source attribution and factuality scores to facilitate successful human-LLM collaboration.

\bibliographystyle{ACM-Reference-Format}
\bibliography{100-reference}

\end{document}